# TAXATION AS AN INSTRUMENT OF STIMULATION OF INNOVATION-ACTIVE BUSINESS ENTITIES

Андрей Сергеевич Нечаев

Andrej Sergeevich Nechaev

Summary: The analysis of the theoretical material revealed the lack of consensus on definition of the tax stimulation of innovation-active business entities within the regional taxation. The definition «tax stimulation of innovation-active business entities» is specified. The author's view on this definition within the regional taxation is submitted. The correlation between the regional taxation and the indicators of financial and property condition of the business entities is considered. The algorithm of the tax rate of corporate property tax, motor vehicle tax and income tax transferred to the regional budget as a function of the rate of return on assets is developed. The innovative method for the calculation of the indicators within the regional taxation on the property tax, vehicle tax and income tax, which is transferred to the budget of the administrative subjects of the Russian Federation is developed. This method allows reducing tax burden for the enterprises with a correction to the coefficient of a profitability of fixed assets.

Innovation – is a highly complex organizational and economic process, based on the use of two kinds of potentials: research potential related to the latest technology and equipment, and the intellectual potential associated with an ability of management to innovate at all stages of production and business operations. The most important element of this process is finding and managing the significant financial resources [1, 2].

There were discussed both general and specific problems of innovative growth, creating conditions for comprehensive modernization and mechanisms of innovative development of regions in the debate at the II Forum of Russian regions in February 2011. The deputy Minister of Economic Development of the Russian Federation has particularized on the draft strategy of innovative development of the Russian Federa-tion until 2020 prepared by the Russian Ministry of Economic Development. The special role is assigned to innovation in regions.

Subject to the foregoing, we can conclude that one of the most important instruments to stimulate innovation businesses at the regional level is tax incentives.

The scrutiny of the current situation, trends and regularities of the innovation sphere development suggests the different nature of an effect of tax instruments for innovation activity enhancing [3, 4]. At the present time there are various forms of tax incentives for businesses provided by the tax laws of the Russian Federation [5]. Among them we can highlight the tax incentives, tax deduction, investment tax credit, exemption from taxation (tax holidays), deferred taxation [6]. However, these forms do not take into account how efficiently basic production assets are used in business entities that in the final analysis affect the financial and property status of an enterprise as a whole.

At the same time the tax rates within the regional taxation are set fixed for the duration of fixed assets using. Thus, it does not encourage enterprises to the most rational using of available resources and investing to the innovative development.

The analysis of the theoretical material [7, 8, 9] revealed the lack of consensus on definition of the tax stimulation of innovation-active business entities within the regional taxation.

Given the above, we should specify the definition of the tax stimulation of innovation-active business entities.

The tax stimulation of innovation-active business entities is a set of actions aimed at the intensification of the interest of innovation-active business entities in the efficient using the fixed assets in the social useful activities by the prospect of obtaining the tax incentives as an instrument to stimulate innovative development. In this case, the tax stimulation of innovation-active business entities within the regional taxation is to adjust the tax rate of the property tax, vehicle tax and income tax which is transferred to the budgets of administrative subjects of the Russian Federation, by a factor reflecting a property and financial status of a developing innovative enterprise.

Next, we represent the algorithm of tax rates adjustments within the regional taxation (the property tax, vehicle tax, income tax which is transferred to the budgets of the administrative subjects of the Russian Federation) in the corresponding i-th period for innovation-active business entities.

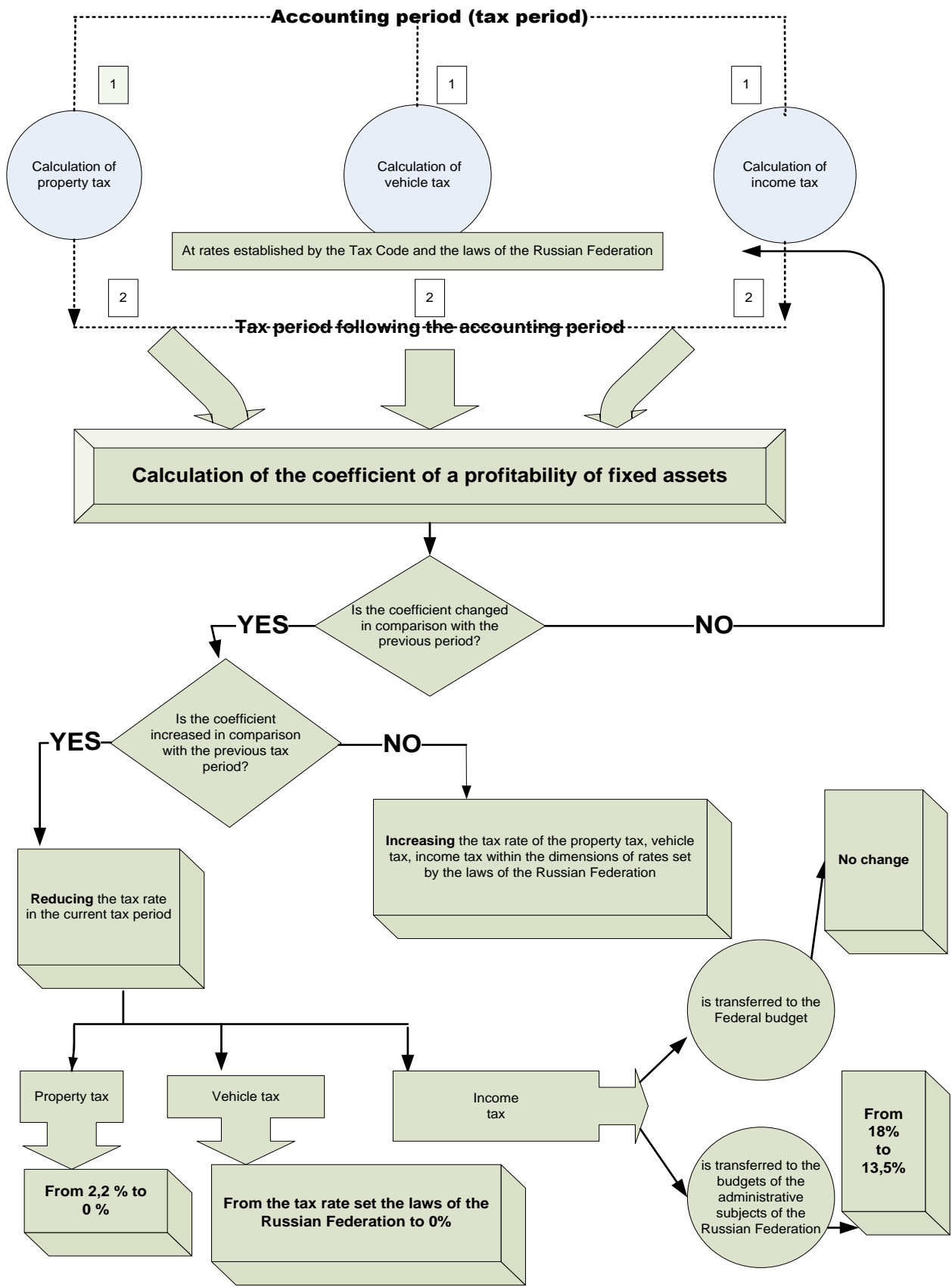

Figure 1. – Algorithm of tax rates adjustments within the regional taxation in the corresponding i-th period for innovation-active business entities

Figure 1 graphically illustrates the process of tax rates adjustments of the property tax, vehicle tax and income tax, which is transferred to the budget of the administrative subjects of the Russian Federation. In the accounting tax period, pointed by number 1, calculation of the relevant taxes with current tax rates within the regional taxation is specified by the Tax Code and laws of administrative subjects of the Russian Federation. In the tax period following the accounting period, pointed by number 2, an enterprise calculates the coefficient of a profitability of fixed assets by means of the balance sheet profit and residual value of fixed asset. Depending on the reducing or increasing of the coefficient of a profitability of fixed assets tax rates of the property tax, vehicle tax and income tax, which is transferred to the budget of the administrative subjects of the Russian Federation, are changed according to the methods below.

Thus, innovation-active business entities annually recalculates the tax rates of the property tax, vehicle tax and income tax, which is transferred to the budget of the administrative subjects of the Russian Federation, depending on the coefficient of a profitability of fixed assets.

It should be emphasized that the development of innovative activity, transition to innovation-based economics – is the only method of reconstruction and development of Russian economic potential.

The method of the tax rates adjustments of the property tax, vehicle tax and income tax, which is transferred to the budget of the administrative subjects of the Russian Federation, by the coefficient of a profitability of fixed assets is represented below.

Designations, used for the detailing this method, are presented in the table 1.

Table 1

Enterprise's economic indices, used for tax rates adjustments of taxes within the regional taxation

| # | Index | Designations |
|---|-------|--------------|
| 1 | Year | i |
| 2 | Amount of depreciation | A |
| 3 | Useful period of the fixed assets | N |
| 4 | Historical cost of fixed assets | F |
| 5 | Percent of depreciation in the linear depreciation method without acceleration factor (in fractions) | $\frac{1}{N}$ |
| 6 | Residual value of fixed assets | R |
| 7 | Average annual cost of fixed assets | M |
| 8 | Proceeds | E |
| 9 | Balance sheet profit | P |
| 10 | Tax base for the vehicle tax (engine power of a vehicle in horsepower; the trust of a jet engine; a gross tonnage; unit of a vehicle) | Bv |
| 11 | The property tax rate (in fractions) | t |
| 12 | The vehicle tax rate (in fractions) | v |
| 13 | The income tax rate (in fractions) | y |
| 14 | The income tax rate transferred to the federal budget | fy |
| 15 | The income tax rate transferred to the regional budget | ry |
| 16 | The coefficient of a profitability of fixed assets | PrA |
| 17 | The coefficient of turnover profitability | PrT |
| 18 | The sum of the property tax | St |
| 19 | The sum of the vehicle tax | Sv |
| 20 | The sum of the income tax | Sp |
| 21 | The sum of the income tax transferred to the federal budget | FSp |
| 22 | The sum of the income tax transferred to the regional budget | RSp |
| 23 | The saved sum of the income tax | ΔSp |

Then create the innovative method for the calculation of the indicators within the regional taxation on the property tax, vehicle tax and income tax, which is transferred to the budget of the administrative subjects of the Russian Federation. This method allows reducing tax burden for the enterprises with a correction to the coefficient of a profitability of fixed assets.

The annual amount of an accrued depreciation is calculated using the following formula 1:

$$A = \frac{F}{N}, \text{где } i = 1,2,3,...,N \qquad (1)$$

For the first year of operation a residual value of fixed assets is equal to their historical cost. For the second year a residual value is equal to the difference between the historical cost of fixed assets and the amount of an accumulated depreciation, etc. Therefore, the residual value of fixed assets is calculated using the formula 2:

$$R = F(1 - \frac{i}{N}) \qquad (2)$$

The average annual cost of fixed assets is calculated using equation 3:

$$M = \frac{F}{N}(N - i + 0,5) \qquad (3)$$

Next, calculate the tax property by the formula 4:

$$St_i = t \cdot \frac{F(N - i + 0,5)}{N}$$

$$St = t \cdot \frac{F \cdot N}{2} \qquad (4)$$

According to the provisions of the chapter 28 of the Tax Code of the Russian Federation [10] the vehicle tax is calculated by the formula 5:

$$Sv = v \cdot Bv \qquad (5)$$

Under the chapter 25 of the Tax Code of the Russian Federation the income tax shall be calculated according to the formula 6:

$$Sp = y \cdot P \qquad (6)$$

The coefficient of a profitability of fixed assets is calculated with the equation 7:

$$\Pr A = \frac{P}{M}$$

$$\Pr A_i = \frac{P_i \cdot N}{F(N - i + 0,5)} \qquad (7)$$

Then the property tax rate adjusted for the coefficient of a profitability of fixed assets is calculated as the product of the ratio of the coefficient of a profitability of fixed assets of a previous period to the coefficient of the accounting period and the property tax rate of a previous period by the formula 8:

$$t_i = t_{i-1} \cdot \frac{P_{i-1}}{P_i} \cdot \frac{N - i + 0,5}{N - i + 1,5} \qquad (8)$$

From the next tax period the sum of the property tax depending on the effectiveness fixed assets using with adjustments for the coefficient of a profitability of fixed assets is calculated by the formula 9:

$$St_i = t_{i-1} \cdot \frac{P_{i-1}}{P_i} \cdot \frac{F}{N} \cdot \frac{(N-i+0,5)^2}{N-i+1,5} \qquad (9)$$

The sum of the property tax adjusted for the coefficient of a profitability of fixed assets is calculated by the equation 10:

$$S_t = \frac{F}{N}\left(t_1(N-0,5) + \sum_{i=2}^{N} \frac{t_{i-1} \cdot P_{i-1} \cdot (N-i+0,5)^2}{P_i \cdot (N-i+1,5)}\right) \qquad (10)$$

In accordance with the chapter 28 of the Tax Code of the Russian Federation, legislative (representative) bodies of the administrative subjects of the Russian Federation determine the vehicle tax rate within the limits established by the Tax Code. This tax rate is calculated by the formula 5.

Then the vehicle tax rate, adjusted for the coefficient of a profitability of fixed assets, is calculated as the product of the ratio of the coefficient of a profitability of fixed assets of the previous period to the coefficient of the accounting tax period and the vehicle tax rate of the previous tax period by the formula 11:

$$v_i = v_{i-1} \cdot \frac{P_{i-1}}{P_i} \cdot \frac{N-i+0,5}{N-i+1,5} \qquad (11)$$

From the next tax period the sum of the vehicle tax adjusted for the coefficient of a profitability of fixed assets is calculated by the formula 12:

$$Sv_i = v_{i-1} \cdot \frac{P_{i-1}}{P_i} \cdot \frac{N-i+0,5}{N-i+1,5} \cdot Bv \qquad (12)$$

The sum of the vehicle tax with adjusted for the coefficient of a profitability of fixed assets is calculated by the formula 13:

$$Sv = Bv\left(v_1 + \sum_{i=2}^{N} \frac{P_{i-1}}{P_i} \cdot \frac{(N-i+0,5) \cdot v_{i-1}}{N-i+1,5}\right) \qquad (13)$$

In accordance with the Article # 284 of the Tax Code of the Russian Federation the income tax rate is determined at 20 percents. The tax's sum, calculated by tax rate at 2 percents, is transferred to the federal budget, and the tax's sum, calculated by tax rate at 18 percents, is transferred to budgets of the administrative subjects of the Rus-sian Federation. The rate on the income tax, which is transferred to the budgets of the administrative subjects of the Russian Federation, could be reduced for certain categories of tax payers to 13,5 percents by the laws of subjects of the Russian Federation. Consequently, if the method developed by us is applied in the income tax, the income tax rate could be reduced to the lowest possible value which is determined at 13,5 percents by the Tax Code to transfer to administrative subjects of the Russian Federation.

In accordance with the chapter 25 of the Tax Code of the Russian Federation the income tax is calculated by the formula 6. Then tax's sums transferred to the federal and regional budgets is calculated by the formulas 15 and 16 respectively:

$$FSp = fy \cdot P \qquad (15)$$

$$RSp = ry \cdot P \qquad (16)$$

For all that: $FSp = 0,02$; $13,5 \leq RSp \leq 18$.

Therefore the income tax rate transferred to the regional budget adjusted for the coefficient of a profitability of fixed assets could be reduced from 18 percents to 13,5 percents and calculated as a product of ratio of the coefficient of a profitability of fixed assets of the previous period to the coefficient of the accounting tax period and the income tax rate of the previous tax period which is transferred to the regional budget by the formula 17:

$$ry_i = ry_{i-1} \cdot \frac{P_{i-1}}{P_i} \cdot \frac{N - i + 0,5}{N - i + 1,5} \qquad (17)$$

The sum of the income tax transferred to the regional budget from the next tax period with adjustments for the coefficient of a profitability of fixed assets would be calculated by formula 18:

$$RSp_i = ry_{i-1} \cdot P_{i-1} \cdot \frac{N - i + 0,5}{N - i + 1,5} \qquad (18)$$

For all that in frames of the current tax legislation $RSp_i$ could be any value in following limits $[0,135 \cdot Pi; 0,18 \cdot Pi]$.

The sum of the income tax transferred to the regional budget with adjustments for the coefficient of a profitability of fixed assets is calculated by formula 19:

$$RSp = 0{,}18P_1 + \sum_{i=2}^{N} \frac{ry_{i-1} \cdot P_{i-1}(N-i+0{,}5)}{N-i+1{,}5} \qquad (19)$$

For all that $RSp$ could be any value in limits determined by the current legislation $\left[0{,}135\sum_{i=1}^{N} Pi; 0{,}18\sum_{i=1}^{N} Pi\right]$.

Then the sum of the income tax is calculated by formula 20:

$$Sp = 0{,}2P_1 + 0{,}02\sum_{i=2}^{N} P_i + \sum_{i=2}^{N} \frac{ry_{i-1} \cdot P_{i-1}(N-i+0{,}5)}{N-i+1{,}5} \qquad (20)$$

For all that $Sp$ could be any value in limits determined by the current legislation $\left[0{,}155\sum_{i=1}^{N} Pi; 0{,}2\sum_{i=1}^{N} Pi\right]$.

Thus, we has developed the method of adjustments of the tax rates on the property tax, vehicle tax and profits tax transferred to the budget of the administrative subjects of the Russian Federation, for the coefficient of a profitability of fixed assets. This method is aimed to facilitate the innovative development of enterprises on regional level, reducing tax burden for innovation-active businesses entities and thereby releasing funds invested in innovative component of the strategic development of business entities.